%% file: main.tex
\documentclass[sigconf,screen]{acmart}

\AtBeginDocument{%
  }

\copyrightyear{2026}
\acmYear{2026}
\setcopyright{cc}
\setcctype{by}
\acmConference[IDE '26]{3rd International Workshop on Integrated Development Environments }{April 12--18, 2026}{Rio de Janeiro, Brazil}
\acmBooktitle{3rd International Workshop on Integrated Development Environments (IDE '26), April 12--18, 2026, Rio de Janeiro, Brazil}
\acmDOI{10.1145/3786151.3788599}
\acmISBN{979-8-4007-2384-1/2026/04}

\input{preamble}
\input{commands}

\begin{document}

\title{From Detection to Prevention: Explaining Security-Critical Code to Avoid Vulnerabilities}

\author{Ranjith Krishnamurthy}
\orcid{0000-0002-0906-5463}
\affiliation{%
  \institution{Paderborn University \& Fraunhofer IEM}
  \city{Paderborn}
  \country{Germany}}
\email{ranjithk@hni.uni-paderborn.de}

\author{Oshando Johnson}
\orcid{0009-0001-1884-7969}
\affiliation{%
  \institution{Fraunhofer IEM}
  \city{Paderborn}
  \country{Germany}}
\email{oshando.johnson@iem.fraunhofer.de}

\author{Goran Piskachev}
\orcid{0000-0003-4424-5838}
\affiliation{%
  \institution{Amazon Web Services}
  \city{Berlin}
  \country{Germany}}
\email{gpiskach@amazon.com}
\authornote{The content of this paper does not reflect the author’s position at Amazon Web Services.}

\author{Eric Bodden}
\orcid{0000-0003-3470-3647}
\affiliation{%
 \institution{Paderborn University \& Fraunhofer IEM}
 \city{Paderborn}
 \country{Germany}}
\email{eric.bodden@uni-paderborn.de}

\renewcommand{\shortauthors}{Krishnamurthy et al.}

\begin{abstract}
Security vulnerabilities often arise unintentionally during development due to a lack of security expertise and code complexity. Traditional tools, such as static and dynamic analysis, detect vulnerabilities only after they are introduced in code, leading to costly remediation. This work explores a proactive strategy to prevent vulnerabilities by highlighting code regions that implement security-critical functionality---such as data access, authentication, and input handling---and providing guidance for their secure implementation. We present an IntelliJ IDEA plugin prototype that uses code-level software metrics to identify potentially security-critical methods and large language models (LLMs) to generate prevention-oriented explanations. Our initial evaluation on the Spring-PetClinic application shows that the selected metrics identify most known security-critical methods, while an LLM provides actionable, prevention-focused insights. Although these metrics capture structural properties rather than semantic aspects of security, this work lays the foundation for code-level security-aware metrics and enhanced explanations.
\end{abstract}




\maketitle

\input{content/introcuction}
\input{content/approach}
\input{content/findings}
\input{content/future_work}
\input{content/conclusion}

\section{Acknowledgments}
This research was partially funded by the European Union and the State of North Rhine-Westphalia under the EFRE/JTF NRW 2021–2027 program (grant EFRE-20800510, CyberResilience.nrw).

\bibliographystyle{ACM-Reference-Format}
\bibliography{sample-base}

\end{document}

%% file: preamble.tex
\usepackage{xcolor} 
\usepackage[nolist]{acronym}
\usepackage{amsfonts} 	

\usepackage{subcaption}
\usepackage{xspace}
\usepackage{multirow, makecell}
\usepackage{listings}
\usepackage{soul}

\usepackage{pifont}   

\usepackage{cleveref}
\newcommand\myshade{85}
\colorlet{mylinkcolor}{violet}
\colorlet{mycitecolor}{orange}
\colorlet{myurlcolor}{gray}

\hypersetup{
    linkcolor  = mylinkcolor!\myshade!black,
    citecolor  = mycitecolor!\myshade!black,
    urlcolor   = myurlcolor!\myshade!black,
    colorlinks = true,
  }%

\usepackage{fancybox}   

\definecolor{promptbar}{RGB}{0,0,0}      
\definecolor{promptbg}{RGB}{255,255,255} 

\newcommand{\SystemPromptBox}[3]{%
  \par\medskip\noindent
  \begingroup
  \setlength{\fboxsep}{4pt}%
  \begin{Sbox}%
    \begin{minipage}{0.949\linewidth}%
      \vspace*{-6pt}
      \hspace{-7.2pt}
      \colorbox{promptbar}{%
        \parbox{\dimexpr\linewidth+0.9\fboxsep\relax}{%
          \hspace{4pt}\textcolor{white}{\bfseries #1}%
        }%
      }%
      \par\medskip
      {\noindent #2\par}%
      \par\vspace{6pt}
      {\noindent #3\par}%
    \end{minipage}%
  \end{Sbox}%
  \doublebox{\TheSbox}%
  \endgroup
  \par\medskip
}

\newcommand{\UserPromptBox}[4]{%
  \par\medskip\noindent
  \begingroup
  \setlength{\fboxsep}{4pt}%
  \begin{Sbox}%
    \begin{minipage}{0.949\linewidth}%
      \vspace*{-6pt}
      \hspace{-7.2pt}
      \colorbox{promptbar}{%
        \parbox{\dimexpr\linewidth+0.9\fboxsep\relax}{%
          \hspace{4pt}\textcolor{white}{\bfseries #1}%
        }%
      }%
      \par\medskip
      {\noindent #2\par}%
      \par\vspace{6pt}
      {\noindent #3\par}%
      \par\vspace{6pt}
      {\noindent #4\par}%
    \end{minipage}%
  \end{Sbox}%
  \doublebox{\TheSbox}%
  \endgroup
  \par\medskip
}

%% file: commands.tex
\newcommand{\funRef}[1]{\texttt{\sloppy{#1}}}

\definecolor{Low}{HTML}{E7B416}    
\definecolor{Medium}{HTML}{DB7B2B} 
\definecolor{High}{HTML}{CC3232}   

\newcommand{\status}[3]{{\color{#1}\scalebox{#2}{\ding{#3}}}}

\newcommand{\low}{\status{Low}{0.8}{116}}
\newcommand{\medium}{\status{Medium}{0.8}{108}}
\newcommand{\high}{\status{High}{0.8}{115}}



%% file: content/introcuction.tex
\section{Introduction}
\label{sec:intro}

Modern software systems are rapidly increasing in size and complexity. As a result, security has become a critical concern, as even small coding mistakes can lead to severe vulnerabilities. To address this, developers use Static and Dynamic Application Security Testing (SAST \& DAST)~\cite{elder2022really} for vulnerability detection. While effective at identifying flaws, these tools are reactive, detecting vulnerabilities only after they are introduced into the codebase. Consequently, issues are often discovered late in the development cycle, when remediation is far more costly. These tools are also resource-intensive, requiring significant time and computational effort~\cite{piskachev2021secucheck}.

Many vulnerabilities arise unintentionally due to limited security expertise, high code complexity, and insufficient understanding of the code~\cite{votipka2020understanding}. One approach to address these challenges is to proactively notify developers of security-critical code regions---segments whose operations, if implemented incorrectly, can compromise the system's confidentiality, integrity, or availability. These code regions often handle sensitive data, access databases, and similar security-relevant operations. Since most developers are not security experts, providing explanations of these regions in natural language can make the information more intuitive and actionable. Large Language Models (LLMs) are well-suited for this task, as they can generate context-aware, human-readable explanations tailored to specific needs. Recent work, such as SAFE~\cite{johnson2025explaining}, uses LLMs to explain detected vulnerabilities, whereas we aim to support prevention by proactively explaining security-critical code regions.

In this paper, we present a prototype IntelliJ IDEA plugin that assesses the security-criticality of methods using code-level software metrics and utilizes LLMs to explain their security relevance and suggest preventive measures. The plugin provides proactive, prevention-oriented guidance directly in the IDE, helping developers understand risky code regions and secure coding practices. An initial evaluation on the Spring-PetClinic~\cite{petclinic_java} application confirms feasibility but also reveals that software metrics primarily capture structural properties, such as complexity, rather than security semantics. These insights motivate the development of security-aware metrics that more accurately estimate security-critical code regions and enable more precise and trustworthy LLM-based explanations.

We present the details of our approach and prototype implementation in Section~\ref{sec:approach}. In Section~\ref{sec:findings}, we present the preliminary findings and future directions in Section~\ref{sec:future_work}. Finally, we conclude in Section~\ref{sec:conclusion}. A demonstration of our prototype plugin is available online~\cite{our-demo}.

%% file: content/approach.tex
\section{Approach}
\label{sec:approach}
\begin{figure}[h] 
    \centering
    \includegraphics[scale=0.90]{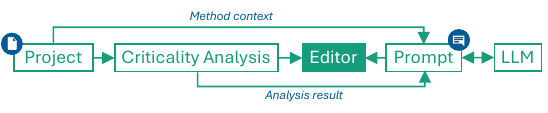}
    \caption{Proposed approach for assessing and explaining security-critical methods.}
    \Description{Proposed approach for assessing and explaining security-critical methods.}
    \label{fig:architecture}
    \vspace{-1.0em}
\end{figure}

In this section, we describe the proposed approach for providing proactive assistance to developers. Figure~\ref{fig:architecture} illustrates the overall workflow. The first stage is the security-criticality assessment, which analyzes the given project to estimate the criticality of all methods. This stage must be fast for proactive assistance. The second stage is explanation generation, which uses the results of the criticality analysis and contextual elements, such as the method body, to create a prompt for generating prevention-oriented explanations. In the final stage, the generated explanations and criticality results are displayed directly in the editor as part of the developer’s workflow. Our prototype is built on the IntelliJ IDEA platform. The following subsections describe criticality assessment (\ref{subsec:criticality_assessment}), explanation generation (\ref{subsec:explanation_gen}), and IDE integration (\ref{subsec:ide_plugin}) stages.

\input{content/criticality_assessment}
\input{content/explanation_generation}
\input{content/plugin_features}

%% file: content/criticality_assessment.tex
\subsection{Security Criticality Assessment}
\label{subsec:criticality_assessment}

The first stage identifies security-critical methods, which are code segments whose incorrect implementation could compromise confidentiality, integrity, or availability. Since our goal is proactive support, this stage must provide deterministic results fast. Several approaches can be used for this task. For instance, static analysis techniques such as taint analysis~\cite{piskachev2021secucheck} can deterministically highlight security-relevant code but are computationally expensive and unsuitable for proactive use. Furthermore, LLM-based criticality assessment can capture semantic aspects of security but introduces latency, suffers from non-determinism, and may hallucinate, limiting its reliability for this task.

As an alternative, we consider metrics-based approaches. Although ISO/IEC 5055:2021~\cite{iso5055} defines security-related quality measures, such measures often rely on more complex analyses. Therefore, for proactive use, we rely on efficiently computable lightweight code-level metrics. Prior work shows that software metrics such as cyclomatic complexity (CC), lines of code (LOC), and lack of cohesion of methods (LCOM) can serve as indirect indicators of security relevance~\cite{chowdhury2011using, siavvas2021hierarchical}. Our prototype currently supports these three metrics, allowing the user to select one based on their analysis needs. CC has been shown to correlate with defect-proneness~\cite{chowdhury2011using} and can highlight complex methods that are prone to security risks. LOC reflects method size, as larger methods are more complex to reason about and may conceal vulnerabilities. LCOM captures cohesion, where high values often indicate unclear responsibilities and potential misuse of data. While these metrics do not directly capture security semantics, they provide a fast and deterministic proxy for identifying potentially security-critical regions.

We use the CK tool~\cite{ck}, which statically computes various code-level metrics in Java projects. Higher metric values (e.g., greater CC or LOC) generally indicate greater risk, while zero suggests non-criticality. Therefore, methods with zero values are filtered out, and the rest are sorted in descending order and grouped into High, Medium, and Low levels using quantile-based equal-frequency binning~\cite{liu2002discretization}, which balances these levels, avoids skewing, and makes the results comparable across metrics within a project. These results form the basis for the subsequent explanation generation stage.

%% file: content/explanation_generation.tex
\subsection{LLM-based Explanation}
\label{subsec:explanation_gen}

To obtain LLM-based security criticality explanations, we use a zero-shot role-playing prompt consisting of two parts: a system and a user prompt. The system prompt defines the LLM’s role as an expert in security-criticality and outlines its tasks. The user prompt provides the method body, the metric with its value, and requests a brief security-focused explanation with precautionary steps. This prompt ensures that explanations remain tied to the given metric, emphasize prevention, and avoid overly generic responses. We use the following system and user prompts for the prototype.

\vspace{1pt}
\SystemPromptBox
  {\textsc{System Prompt}}
  {You are an assistant with expertise in explaining the security criticality of
   software in regard to the system’s confidentiality, integrity, and availability.
   For a given code snippet, you will be provided with the name of a software metric
   that measures its criticality, the interpretation of the metric value, and the
   value of the metric. One of the following metrics will be provided: cyclomatic
   complexity, lines of code, or lack of cohesion of methods. Your task is to explain
   why the code snippet is security critical using the provided metric and give steps
   to prevent possible security exploits due to mistakes in the code snippet.}
  {\texttt{<Guidelines and response format for LLM>}}
\vspace{-6pt}

\UserPromptBox
  {\textsc{User Prompt}}
  {Explain why the code snippet is security critical based on the metric and provide concise steps to prevent security exploits.}
  {Code Snippet: \texttt{<method body>}}
  {Metric: \texttt{<metric name (it's value interpretation): metric value>}}

\vspace{1pt}
We use OpenAI models via the Azure OpenAI API, primarily GPT-5 (gpt-5-2025-08-07), the latest reasoning-capable model at the time of this paper. Additionally, we experiment with GPT-4 (gpt-4o-2024-08-06) to assess the impact of model choice on explanation quality and efficiency. These models were selected for their popularity and strong performance on code-understanding tasks. Benchmarking across LLMs is out of scope, as our focus is on evaluating the prototype’s effectiveness.

\begin{figure*}[h]
    \centering
    \includegraphics[width=\textwidth]{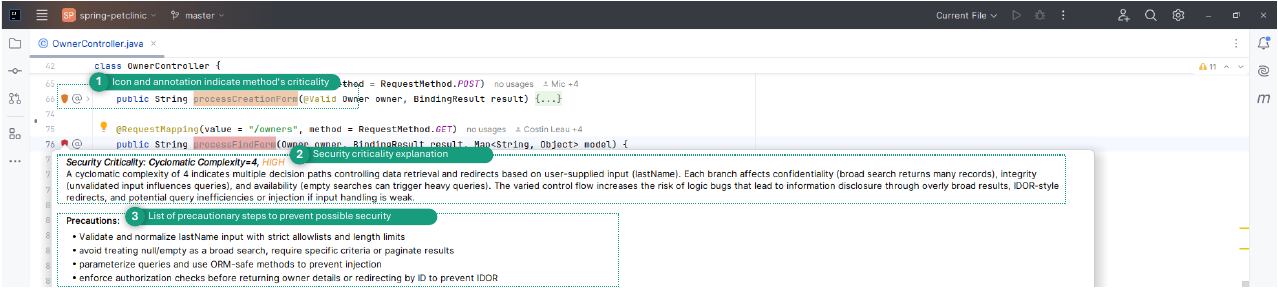}
    \caption{Screenshot of the IntelliJ plugin showing icons and annotations for security-critical methods (\includegraphics[width=0.014\textwidth]{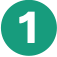}), security-criticality explanations (\includegraphics[width=0.014\textwidth]{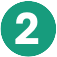}), and precautionary steps (\includegraphics[width=0.014\textwidth]{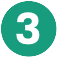}) using GPT-5 to prevent introducing vulnerabilities.}
    \Description{Screenshot of the IntelliJ plugin showing icons and annotations for security-critical methods (\includegraphics[width=0.014\textwidth]{figures/marker1.png}), security-criticality explanations (\includegraphics[width=0.014\textwidth]{figures/marker2.png}), and precautionary steps (\includegraphics[width=0.014\textwidth]{figures/marker3.png}) using GPT-5 to prevent introducing vulnerabilities.}
    \label{fig:plugin-screenshot}
    \vspace{-0.4em}
\end{figure*}

%% file: content/plugin_features.tex
\subsection{IDE Integration}
\label{subsec:ide_plugin}

Based on the results of the criticality analysis (Sub-Section~\ref{subsec:criticality_assessment}), the plugin annotates security-critical methods with icons and color cues indicating their criticality levels (Figure~\ref{fig:plugin-screenshot}). By default, it prioritizes High and Medium methods, while Low-criticality methods can be shown on demand to reduce visual clutter. These visual cues align with developers’ workflows, supporting rapid triage and prioritization during coding and review while reducing the time required to interpret assessment results. Hovering over the gutter icons or annotations applied to a method name (\includegraphics[width=0.014\textwidth]{figures/marker1.png}) displays a tooltip showing the metric value, its criticality level, and an explanation generated by an LLM (Sub-Section~\ref{subsec:explanation_gen}). This explanation describes why the method is security-critical (\includegraphics[width=0.014\textwidth]{figures/marker2.png}) and suggests precautionary steps (\includegraphics[width=0.014\textwidth]{figures/marker3.png}) to prevent potential vulnerabilities. Additionally, plugin actions are accessible from the editor’s context menu, allowing users to select among the supported software metrics and configure settings.

%% file: content/findings.tex
\section{Evaluation}
\label{sec:findings}

To evaluate our prototype, we ran the IntelliJ plugin on the vulnerable Spring-PetClinic~\cite{petclinic_java} application, which contains 99 methods. For GPT-5, which operates with a fixed temperature of 1, we set the reasoning level to minimal to reduce processing time. For GPT-4, we set the temperature to 0 to ensure reproducible outputs. Experiments were conducted on a Windows 11 machine with an AMD Ryzen 7 CPU and 16 GB RAM. In the following subsection, we present the preliminary findings observed from these experiments.

\subsection{Preliminary Findings}
\label{subsec:prelim_findings}

We organize our findings around three key aspects: runtime performance, explanation quality, and the effectiveness of the criticality assessment based on software metrics. 

\textit{Runtime:} Metric computation for all 99 methods completes in about 2 seconds on average. Explanation generation is slower, averaging $\approx$7.2 seconds per method using GPT-5 and $\approx$1.3 seconds using GPT-4. This latency in LLMs can delay proactive assistance, so the plugin immediately displays the computed metric and criticality level with a placeholder message: “Generating explanation, please wait”. Explanations are then generated by criticality order and shown incrementally as they become available, providing early visibility into critical regions and timely, actionable insights.

\textit{Explanation quality:} GPT-4 generates explanations faster but often produces generic descriptions that overlook code-specific security risks. In contrast, GPT-5 provides more precise, context-aware explanations that highlight relevant operations, though at higher latency and with some generalized recommendations. This reflects a trade-off between explanation depth and efficiency that future work should aim to balance.

\textit{Criticality Assessment:} The distribution of criticality levels varies across metrics: LOC identifies 35 methods as High, 55 as Medium, and 9 as Low; CC flags 5 High, 13 Medium, and 81 Low; and LCOM classifies 23 High, 26 Medium, and 21 Low. To assess detection accuracy, Table~\ref{tab:comparison} compares our results with the ground truth from prior work~\cite{piskachev2021secucheck, krishnamurthy2022extent}, which identified eight vulnerable methods. The numbers in the table represent each method’s rank within the project based on our criticality assessment (rank 1 = most critical). Our approach flags all eight known vulnerable methods as High~\textsuperscript{\high} or Medium~\textsuperscript{\medium} using LOC, three as High or Medium using CC, and one as Medium using LCOM. These results indicate that software metrics, particularly LOC and CC, show potential for identifying vulnerable methods as security-critical in our evaluation.

\begin{table}[h]
\centering
\renewcommand{\arraystretch}{1.2}

\begin{tabular}{|p{4.2cm}|c|c|c|}
\hline
\multirow{2}{4.2cm}{\centering\textbf{Method}} & \multicolumn{3}{c|}{\textbf{Criticality Rank by}} \\ \cline{2-4}

 & \textbf{CC} & \textbf{LOC} & \textbf{LCOM} \\ \hline

\centering \makecell{\funRef{OC.showOwner}} & 50~\low & 27~\high & 56~\low \\ \hline
\centering \makecell{\funRef{OC.initUpdateOwnerForm}} & 53~\low & 29~\high & 60~\low \\ \hline
\centering \makecell{\funRef{OC.processFindForm}} & 04~\high & 01~\high & 58~\low \\ \hline
\centering \makecell{\funRef{ORCI.findOwner}} & 60~\low & 55~\medium & 46~\medium \\ \hline
\centering \makecell{\funRef{ORCI.findById}} & 27~\low & 24~\high & -- \\ \hline
\centering \makecell{\funRef{ORCI.findByLastName}} & 28~\low & 25~\high & -- \\ \hline
\centering \makecell{\funRef{ORCI.save}} & 08~\medium & 12~\high & -- \\ \hline
\centering \makecell{\funRef{PC.processCreationForm}} & 11~\medium & 10~\high & 61~\low \\ \hline

\end{tabular}

\vspace{0.4em}
\begin{minipage}{\linewidth}
\centering
{\fontsize{6.4}{8}\selectfont
OC = OwnerController, PC = PetController, ORCI = OwnerRepositoryCustomImpl\\
High~\textsuperscript{\high}, Medium~\textsuperscript{\medium}, Low~\textsuperscript{\low}
}
\end{minipage}

\vspace{0.4em}
\caption{Software metrics-based criticality assessment of vulnerable methods~\cite{piskachev2021secucheck, krishnamurthy2022extent} from the Spring-PetClinic app.}
\label{tab:comparison}
\vspace{-1.4em}
\end{table}

In addition to the quantitative findings, the evaluation highlights practical advantages of the proposed approach. It supports early security reasoning by rapidly identifying potentially critical code regions and providing prevention-oriented explanations in natural language within the editor, helping developers follow secure coding practices. Furthermore, organizing criticality levels (High, Medium, and Low) using quantile-based equal-frequency binning~\cite{liu2002discretization} transforms raw analysis results into actionable prioritization, enabling developers to focus attention and effort where it matters most. However, since the current prototype relies on software metrics, which are indirect indicators of security, it may produce false positives, flagging non-critical methods as critical. In such cases, the LLM still provides reasonable---yet more generic---preventive guidance based on the method body. This behavior stems from the limitations of software metrics (see Sub-Section~\ref{subsec:limit}) and can be mitigated through the security-aware metrics (see Section~\ref{sec:future_work}). Furthermore, our proposed approach is theoretically language agnostic. While the current implementation targets Java, we also tested the plugin on the Kotlin version of the vulnerable Spring-PetClinic~\cite{krishnamurthy2022extent}, where it exhibited the same behavior, demonstrating promising cross-language potential.

\subsection{Limitations}
\label{subsec:limit}
While the evaluation highlights several strengths of the proposed approach, certain limitations remain in the current prototype. Our observation (Table~\ref{tab:comparison}) shows that software metrics can capture most known vulnerable methods as security-critical, demonstrating their potential. However, they also flag non-critical methods as critical, revealing false positives. Moreover, vulnerable methods are not consistently ranked high—for instance, using LOC and CC, only two appear among the top 10 (Table~\ref{tab:comparison}). This indicates that software metrics alone cannot reliably estimate true security criticality, as they primarily measure structural and maintainability properties rather than security semantics and lack contextual awareness of how methods interact with sensitive data or external components.

A second limitation concerns explanation generation. Although GPT-5 generates more precise, context-specific explanations than GPT-4, it still gives overly generic guidance without pinpointing relevant code locations or variables. This limitation arises because its explanations rely on software metrics that capture structural rather than semantic properties, limiting the depth and accuracy of its insights. Moreover, GPT-5’s general-purpose training may constrain its ability to reason about subtle code-level security semantics.

%% file: content/future_work.tex
\section{Future Work}
\label{sec:future_work}

Future work will focus on improving both the precision of our security-criticality assessment and the contextual depth of the generated explanations. We plan to design code-level security-aware metrics that explicitly capture dangerous operations~\cite{johnson2024detecting} and integrate CWE mappings to connect code regions to their real-world impact. These metrics would not only improve prioritization but also enhance the quality of LLM-generated explanations by providing security semantics for the models to reason over. Furthermore, we also aim to experiment with fine-tuning LLMs to produce richer, context-sensitive explanations and to adapt the level of explanation detail based on the developer’s experience.

We also intend to address the inherent limitations of LLMs, such as hallucination and lack of reproducibility, by introducing validation mechanisms and guardrails similar to AWS Automated Reasoning Checks~\cite{aws_preventing_hallucinations_2024} to enforce reliable, consistent output. Furthermore, we plan to explore integrating our prevention-oriented approach into AI-assisted development workflows. As AI coding agents increasingly shape modern software engineering, embedding security-criticality assessment and explanations as guardrails can enhance the security of AI-generated code. 

%% file: content/conclusion.tex
\section{Conclusion}
\label{sec:conclusion}
In this paper, we presented an IntelliJ plugin that shifts the focus from detecting vulnerabilities to proactively preventing them by highlighting critical code regions and generating explanations using LLMs. Our prototype shows that code-level software metrics enable near-instant feedback, while explanations can be delivered incrementally to support real-time coding. Initial observations confirm the potential of this approach but also reveal limitations---software metrics lack security semantics, and LLMs may produce generic or inconsistent output. These findings open up promising directions for future work, including the development of code-level security-aware metrics, validation of LLM explanations, and integration into both human and AI-assisted development workflows. 

\textit{We envision IDEs integrating code-level security-aware metrics and trustworthy LLM explanations to help developers and AI coding agents understand their code and avoid introducing vulnerabilities, shifting from post-hoc detection to a prevention-first approach.}

%% file: sample-base.bib
@String{Springer = "Springer-Verlag" }

@inproceedings{votipka2020understanding,
  title={{Understanding security mistakes developers make: Qualitative analysis from Build it, Break it, Fix it}},
  author={Votipka, Daniel and Fulton, Kelsey R and Parker, James and Hou, Matthew and Mazurek, Michelle L and Hicks, Michael},
  booktitle={29th USENIX Security Symposium (USENIX Security 20)},
  pages={109--126},
  year={2020}
}

@inproceedings{piskachev2021secucheck,
  title={{SecuCheck: Engineering configurable taint analysis for software developers}},
  author={Piskachev, Goran and Krishnamurthy, Ranjith and Bodden, Eric},
  booktitle={2021 IEEE 21st International Working Conference on Source Code Analysis and Manipulation (SCAM)},
  pages={24--29},
  year={2021},
  organization={IEEE}
}

@article{elder2022really,
  title={{Do I really need all this work to find vulnerabilities? An empirical case study comparing vulnerability detection techniques on a Java application}},
  author={Elder, Sarah and Zahan, Nusrat and Shu, Rui and Metro, Monica and Kozarev, Valeri and Menzies, Tim and Williams, Laurie},
  journal={Empirical Software Engineering},
  volume={27},
  number={6},
  pages={154},
  year={2022},
  publisher={Springer}
}

@article{siavvas2021hierarchical,
  title={{A hierarchical model for quantifying software security based on static analysis alerts and software metrics}},
  author={Siavvas, Miltiadis and Kehagias, Dionysios and Tzovaras, Dimitrios and Gelenbe, Erol},
  journal={Software Quality Journal},
  volume={29},
  number={2},
  pages={431--507},
  year={2021},
  publisher={Springer}
}

@article{liu2002discretization,
  title={{Discretization: An Enabling Technique}},
  author={Liu, Huan and Hussain, Farhad and Tan, Chew Lim and Dash, Manoranjan},
  journal={Data mining and knowledge discovery},
  volume={6},
  number={4},
  pages={393--423},
  year={2002},
  publisher={Springer}
}

@online{ck,
    author = {Aniche, Maurício},
    title = {{Code-level metrics for Java using static analysis}},
    year = {2024},
    url = {https://github.com/mauricioaniche/ck/},
    publisher = {Aniche, Maurício},
    urldate = {2025-09-22},
    type = {GitHub}
}

@online{our-demo,
    author = {Fraunhofer IEM},
    title = {{Proactively explaining security-critical code.}},
    year = {2025},
    url = {https://github.com/fraunhofer-iem/critical-marker-plugin},
    publisher = {Fraunhofer IEM},
    urldate = {2025-09-22},
    type = {GitHub}
}

@online{petclinic_java,
    author = {Contrast Community},
    title = {{Vulnerable Spring PetClinic}},
    year = {2019},
    url = {https://github.com/contrast-community/spring-petclinic},
    publisher = {Contrast Community},
    urldate = {2025-09-22},
    type = {GitHub}
}

@inproceedings{krishnamurthy2022extent,
  title={{To what extent can we analyze Kotlin programs using existing Java taint analysis tools?}},
  author={Krishnamurthy, Ranjith and Piskachev, Goran and Bodden, Eric},
  booktitle={2022 IEEE 22nd International Working Conference on Source Code Analysis and Manipulation (SCAM)},
  pages={230--235},
  year={2022},
  organization={IEEE}
}

@inproceedings{johnson2024detecting,
  title={{Detecting Security-Relevant Methods using Multi-label Machine Learning}},
  author={Johnson, Oshando and Piskachev, Goran and Krishnamurthy, Ranjith and Bodden, Eric},
  booktitle={Proceedings of the 1st ACM/IEEE Workshop on Integrated Development Environments},
  pages={101--106},
  year={2024}
}

@article{chowdhury2011using,
  title={{Using complexity, coupling, and cohesion metrics as early indicators of vulnerabilities}},
  author={Chowdhury, Istehad and Zulkernine, Mohammad},
  journal={Journal of Systems Architecture},
  volume={57},
  number={3},
  pages={294--313},
  year={2011},
  publisher={Elsevier}
}

@online{aws_preventing_hallucinations_2024,
    author = {Antje Barth},
    title = {{Prevent factual errors from LLM hallucinations with mathematically sound Automated Reasoning checks}},
    year = {2024},
    url = {https://aws.amazon.com/blogs/aws/prevent-factual-errors-from-llm-hallucinations-with-mathematically-sound-automated-reasoning-checks-preview/},
    publisher = {Amazon Web Services},
    urldate = {2025-09-22},
    type = {Blog post}
}

@online{iso5055,
    author = {International Organization for Standardization},
    title = {{ISO/IEC 5055:2021 — Automated source code quality measures}},
    year = {2021},
    url = {https://www.iso.org/standard/80623.html},
    publisher = {International Organization for Standardization},
    urldate = {2025-09-22},
}

@inproceedings{johnson2025explaining,
  title={{Explaining Software Vulnerabilities with Large Language Models}},
  author={Johnson, Oshando and Fomina, Alexandra and Krishnamurthy, Ranjith and Chaudhari, Vaibhav and Shanmuganathan, Rohith Kumar and Bodden, Eric},
  booktitle={2025 40th IEEE/ACM International Conference on Automated Software Engineering Workshops (ASEW)},
  pages={194--198},
  year={2025},
  organization={IEEE}
}
